\documentstyle[12pt]{article}
\normalbaselines 
\begin{document}

\begin{center}
\vspace*{1.0cm}

{\LARGE {\bf  Duality and the cosmological constant\footnote{To appear
in International Journal of Theoretical Physiscs, Vol 36, No. 9, (1997) }
\\}}
\vskip 1.5cm

{\large {\bf Hadi Salehi\footnote { e-mail address:
salehi@netware2.ipm.ac.ir}}}

\vskip 0.5 cm

Institute for Studies in Theoretical Physics and
Mathematics, \\ P.O.Box 19395-1795, Niavaran-Tehran, Iran
\\ and\\
Arnold Sommerfeld Institute for Mathematical Physics, TU Clausthal, \\
Leibnizstr. 10, D-38678 Clausthal-Zellerfeld, Federal
Republic of Germany
\end{center}

\vspace{1cm}

\begin{abstract}
A dynamical
theory is studied in which a 
scalar field $\phi$ in Einstein-Minkowski space 
is coupled to the four-velocity $N_{\mu}$ 
of a preferred inertial observer in that space.
As a consistent requirement on this coupling we study 
a principle of duality invariance of the dynamical mass-term of $\phi$
at some universal length in the small-distance regime. In the
large-distance regime
duality breaking can be introduced by giving a back-ground value to 
$\phi$ and a back-ground direction to $N_{\mu}$. 
It is shown that, in an appropriate approximation, duality
breaking can be related to the emergence of a characteristic 
phase in which the condensation of
the ground state allows massive excitations with a
characteristic scale of squared mass which agrees with present observational 
bound for the
cosmological constant. 
\end{abstract}

\vspace{2cm}

\begin{center}  
{\large {\bf  I-Introduction\\ }} 
\end{center}
\vspace{1cm}

In quantum field theory the structure of the vacuum is recognised to be 
interrelated with the condensation of 
scalar fields, a phenomenon characterised by remnant 
constant vacuum expectation values for those fields.  
In general, such scalar (vacuum)
condensates carry informations about various mass scales 
which are characteristic of the dynamical properties of the vacuum.
For example, a particular type of massive particle 
excitation of vacuum can be characterised by the 
mass scale corresponding to the massive excitations around a given scalar
condensate.\\
The condensation of the vacuum and the related appearance of characteristic
mass scales may have an important effect on the energy content of the vacuum,
because one expects that
the cosmological constant receives potential contributions
from any scale of mass which can be extracted from the mass spectrum of
physical fields in quantum field theory. 
In particular, any massive particle excitation around
a given scalar condensate can provide a particular type of
contribution to the total value of the cosmological constant
via the corresponding mass scale. 
In this way
a non-vanishing contribution to the total energy density of vacuum may 
arise from the mass scale of any scalar 
condensate.\\
This kind of contribution, however, meets with an immediate difficulty. 
For example, the isolated contribution of the mass scale of 
the Higgs field in the standard 
model predicts a value for the cosmological constant which is 
many orders of magnitude away from 
the observational bound of that constant. \\ 
A solution of this problem which is a particular manifestation of the
the cosmological constant problem, see [1] and references therein, 
may really require a 
theoretical scheme in which  explicit recognization is given to  
the expected sensitivity of the total
value of the cosmological constant to the entire mass spectrum of
physical fields in quantum field theory, because in a
unified theory the latter fields may not be independent so that 
unexpected 
cancellations among various contributions of the corresponding
mass scales may occur. Such a theoretical scheme is likely to 
reduce the issue of the
cosmological constant to a picture in which  
the consistent contribution to the total value of that constant  
comes from a preferred
mass scale of the vacuum, namely that assigning a cosmological range to the
massive excitations of the vacuum. \\
It must be emphasized that this picture 
is very subtile, and it is extremely
difficult to find out how the contribution of the preferred mass scale
among the other potential contributions could be established. 
But a preliminary task is to focus on the
the nature of this preferred mass scale.
In the present note we
discuss tentative first steps in this direction.

\vspace{1cm}
\begin{center}  
{\large {\bf  II-The broken phase of Lorentz invariance\\ }} 
\end{center}
\vspace{1cm}                                   

We remark that, in an exact Lorentz invariant vacuum, 
the energy density is,
almost by definition, zero. Therefore, one should expect that the 
defining characteristic of a non-vanishing energy density
in vacuum is a principal violation of Lorentz invariance.
The nature of the preferred mass scale in question depends on 
the consistency of such a picture in an essential way.\\
To be more specific we remark that a principal violation
of Lorentz invariance may act in the right way to 
yield a consistent contribution to the cosmological constant 
via the mass scale of 
massive excitations around an associated scalar condensate.
Before we present a model along this line, it is necessary 
to collect some general facts concerning what is expected to be the right way
to think of a principal violation of Lorentz invariance.\\ 
Such a violation of Lorentz invariance should in fact be a consequence
of the still unknown principles underlying the unification of
quantum physics and gravity and is expected to manifest itself 
at some characteristic scale in the ultrashort distance regime, described 
by an absolute scale of length $l_0$.
The understanding of the relation of $l_0$ to the Planck length
is an elusive task for quantum gravity. Here we merely note that
the length 
$l_0$ is expected to act as a sort of universal length that 
determines a lower bound to any scale of 
length probed in a measurement process.\\ 
It should be noticed that the existence of such
a universal length in the small-distance regime 
is in contrast with the universal 
requirement of Lorentz invariance. Indeed, no absolute line of demarcation
between small distances and large ones can be defined without
having a positive definite measure of distances, a feature which is
apparently absent in Einstein-Minkowski space.\\ 
It was pointed out by Blokhintsev [2] that, associating to the 
Einstein-Minkowski
space a time-like 
vector $N_{\mu}$,
the so called internal vector,
it is possible to
distinguish between small distances and large ones 
by taking the positive definite interval
\begin{equation}
R^2= (2N_{\mu}N_{\nu}-\eta_{\mu\nu}) x^{\mu} x^{\nu}, ~~~N_{\mu}N^{\mu}=1
\label{1}\end{equation}
Given such a metric, we may determine the absolute size of a distance 
by comparing $R$ with the universal length $l_0$.     \\
Generally, one would expect the idea of a 
universal length and the related internal vector 
to play a vital role in the ultrashort distance extrapolation 
of physics. Unfortunately, the practical need of this idea
in the development of our present
day physical concepts (to a large extend) disappeared with the achievement
of renormalizable theories,  which explains why that idea has not
attracted wide attention in the particle physics community. \\
Different meanings are assigned in the literature to an internal 
vector, for a review see [3]. We shall adopt the most obvious
interpretation of the vector $N_{\mu}$ and consider
$N_{\mu}$ at each space-time point as characterising 
the four velocity of a preferred inertial observer. 
The assignment of $N_{\mu}$ to the vacuum singles out then
a coordinate system as preferred, namely that in which the preferred 
inertial observer is at rest. Correspondingly, only
a special group of Lorentz transformations can have 
an intrinsic significance,
namely those which leaves the special form $N_{\mu}=(1,0,0,0)$ invariant.
Physically, this special group can be thought of as being composed of those
transformations which move
a physical system without affecting the preferred rest frame.
Thus, if one limits oneself to this special group, 
the internal vector appears as a
universal field which has the same absolute value in vacuum.
In this respect the internal vector can be considered as corresponding to 
a characteristic property of the vacuum in the the broken phase of 
Lorentz invariance.\\  
We shall not comment here on 
the probable form of the preferred rest
frame defined by $N_{\mu}$, 
because the results we want to present depend only
on general considerations\footnote{
For the
conceivable form of such a rest frame in a laboratory
on the earth a proposal is made in [4] where possible experiments to detect 
it are also discussed.}.

\vspace{1cm}            
\begin{center}  
{\large {\bf  III-The Model\\ }} 
\end{center}
\vspace{1cm}

We first study a 
theory which relates
the broken phase of Lorentz invariance 
characterised by an internal vector $N_{\mu}$  
to the condensation
of an associated scalar field. 
Such a 
condensation is organised in such a way as to emerge
from a dynamical coupling of a (real) scalar field $\phi$
with the internal vector $N_{\mu}$.
We then study the massive excitations around the
scalar condensate and relate the corresponding scale of mass to
the present observational bound on the cosmological constant.\\
To arrive at the dynamical coupling of $\phi$ with $N_{\mu}$,
we start with the current
\begin{equation}
J_{\mu}=-\frac{1}{2}\phi\stackrel{\leftrightarrow}{\partial_{\mu}}\phi^{-1}.
\label{2}\end{equation}
It can simply be checked that
\begin{equation}
\partial_{\mu}J^{\mu}=\phi^{-1}(\Box\phi-\phi^{-1}
\partial_{\mu}\phi \partial^{\mu}\phi)
\label{3}\end{equation}
and
\begin{equation}
J_{\mu}J^{\mu}=\phi^{-2}\partial_{\mu}
\phi \partial^{\mu}\phi
\label{4}\end{equation}
are valid. Putting these relations together, we get the identity
\begin{equation}
\Box\phi+\Gamma\{\phi\}\phi=0,~~~\Gamma\{\phi\}= -J_{\mu}J^{\mu}-\partial_{\mu}J^{\mu}
\label{5}\end{equation}
In what follows we shall call $\Gamma\{\phi\}$ the dynamical mass term.\\
It should be emphasised that
the identity (\ref{5}) is a formal consequence of the definition (\ref{2})
and not a dynamical law for $\phi$. From it, however, 
a large class of dynamical
theories can be obtained in the form of a divergence theory 
by making various assumptions about the current
$J^{\mu}$ in the dynamical mass term in (\ref{5}).
For example a simple model theory may be characterised
by requiring
\begin{equation}
\partial_{\mu}J^{\mu}=0.
\label{6}\end{equation}
It leads, as can simply be checked, to 
a cancellation of the dynamical mass term by the field redefinition
$\sigma=\ln{\phi}$. 
However, to allow for a dynamical coupling of $\phi$ with
the internal vector $N_{\mu}$, the dynamical mass term must take a more 
complicated structure.
In this case, we may study a model theory based on a divergence theory of
the type
\begin{equation}
\partial_{\mu}J^{\mu}=N_{\mu}N_{\nu}J^{\mu} J^{\mu}
\label{7}\end{equation}
Given a divergence theory of the type (\ref{7}), 
we get from the identity
(\ref{5}) the
corresponding equation for $\phi$
\begin{equation}
\Box \phi-(J_{\mu}J^{\nu}+N_{\mu}N_{\nu}J^{\mu}J^{\nu}) \phi=0
\label{8}\end{equation}
We remark that the 
dynamical mass-term in (\ref{8}), is invariant under 
both transformations\footnote{Here we treat the scalar field 
as if it were
a dimensionless quantity. In the natural set of units 
a combination of $\phi$ and the universal length $l_{0}$
may be used to get from $\phi$ a field having the typical dimension of an 
inverse distance.}
\begin{equation}
\phi\leftrightarrow\frac{1}{\phi},~~~N_{\mu}\leftrightarrow -N_{\mu}
\label{9}\end{equation}
which can be performed independently.
This indicates an inherent ambiguity in the theory,
because apparently independent configurations composed of $\phi$ and
$N_{\mu}$ become mathematically interchangeable at any physical
scale of mass which can be predicted by the dynamical mass-term. In 
a simple generalisation this ambiguity can 
significantly be avoided by admitting the universal length $l_0$ to enter
the source of the divergence in (\ref{7}). 
We shall study a theory of the type
\begin{equation}
\partial_{\mu}J^{\mu}=N_{\mu}N_{\nu}J^{\mu} J^{\mu}-
l_{0}N_{\mu}N_{\nu}N_{\gamma}J^{\mu}J^{\nu}J^{\gamma}
\label{10}\end{equation}
From the identity
(\ref{4}) we get now the field equation
\begin{equation}
\Box \phi-(J_{\mu}J^{\nu}+N_{\mu}N_{\nu}J^{\mu}J^{\nu}-
l_{0}N_{\mu}N_{\nu}N_{\gamma}J^{\mu}J^{\nu}J^{\gamma}) \phi=0
\label{11}\end{equation}
The interesting point to observe is that the invariance property of the
dynamical mass-term in (\ref{11}) connects now both transformations 
in (\ref{9}).
That is, the dynamical mass term in (\ref{11}) becomes now 
invariant under a duality transformation connecting 
the interchange of $\phi$ to the reciprocal value
$\phi^{-1}$ to a corresponding reversal of the direction of 
the internal vector $N_{\mu}$. 
The emergence of this duality is considered as reflecting the essential 
feature of the brocken phase of Lorentz invariance at length scales 
$\sim l_0$.\\ 
It is not unreasonable to 
expect that 'macroscopic' duality becomes significantly unstable.
For example, at distances larger than 
the universal length $l_{0}$ the average value of the scalar field
is likely to couple with the matter in the universe. Such a coupling
has, among other things, to introduce a duality
breaking to single out a preferred configuration composed of 
the scalar field $\phi$ and the internal vector $N_{\mu}$ throughout
the space. 
Although the nature of such a duality breaking seems to be 
significantly linked with the presence of matter and a non-trivial
gravitational field [5], it is nevertheless instructive to
study its effect on the dynamical theory defined by (\ref{10}) and 
(\ref{11}) in Einstein-Minkowski space.\\
We shall consider the simplest duality breaking, a preferred
back-ground value $\bar{\phi}$ as an average value of $\phi$ taken over large
distances, and correspondingly a preferred back-ground direction
of the internal vector $N_{\mu}$ as an average value $\bar{J_{\mu}}$ 
of the current
$J_{\mu}$ taken over large distances. The essential feature of such a
duality breaking is that it requires (on dimensional grounds) that 
the duality breaking parameters $\bar{\phi}$ and $N_{\mu}$ be
interrelated via a relation
of the type
\begin{equation}
\bar{J_{\mu}}= \frac{\lambda}{l_{0}} N_{\mu}.
\label{13}\end{equation}
where $\lambda$ can depend only on $\bar{\phi}$.
Since we wish to consider the duality breaking as a large distance
effect, the characteristic  
scale of mass defined by the righthand side of (\ref{13}) must be  
related to those scales of lengths 
which are significantly larger than the universal
length $l_{0}$. This is only the case if $\lambda$
is taken as significantly small.
We assume it to be of order of the ratio of the universal length $\l_0$
and the radius of the universe $R$                     
\begin{equation}
\lambda\sim \frac{l_{0}}{R}.
\label{14}\end{equation}
In this way the duality breaking is considered as a cosmological effect. 
Having assumed a duality breaking of this type, 
we proceed now to compute the effective 
form of the dynamical
mass-term in (\ref{11}) for the divergence theory (\ref{10}).\\ 
First, we may linearise the 
quadratic term in $J_{\mu}$ in the source of 
the divergence (\ref{10}) to find the 
approximation
\begin{equation}
\partial_{\mu}J^{\mu}=N_{\mu}N_{\nu}\bar{J^{\mu}} J^{\mu}-
l_{0}N_{\mu}N_{\nu}N_{\gamma}\bar{J^{\mu}}J^{\nu}J^{\gamma}
\label{15}\end{equation}
Correspondingly, the quadratic term in the dynamical 
mass-term of the field equation (\ref{11}) may be linearised in $J_{\mu}$ 
to yield
\begin{equation}
\Box \phi-(\bar{J_{\mu}}J^{\nu}+N_{\mu}N_{\nu}\bar{J^{\mu}}J^{\nu}-
l_{0}N_{\mu}N_{\nu}N_{\gamma}\bar{J^{\mu}}J^{\nu}J^{\gamma}) \phi=0
\label{14a}\end{equation}
Since the background value of $J_{\mu}$ is of the order of $\lambda$,
the dynamical mass-term can effectively be determined to third order 
of $\lambda$. 
To this aim, we truncate the non-linear term in $J_{\mu}$ from the source of 
the divergence (\ref{15}). Computing then 
the remaining linear term by means of (\ref{13}) we get
\begin{equation}
\partial_{\mu}J^{\mu}\simeq \frac{\lambda}{l_{0}} N_{\mu} J^{\mu}
\label{16}\end{equation}
Now, using (\ref{2}), this can be 
written in terms of $\phi$ as
\begin{equation}
\partial_{\mu}J^{\mu}\simeq \frac{\lambda}{l_{0}} N_{\mu} 
\frac{\partial^{\mu}\phi}{\phi}.
\label{17}\end{equation}
The right hand side of this equation can be linearised in $\phi$ by
using in the dominator the background value $\bar{\phi}$ for $\phi$.
An approximate solution of (\ref{17})
which is compatible with (\ref{13}) can then be given
\begin{equation}
J_{\mu}\simeq \frac{\lambda}{l_{0}} \frac{\phi}{\bar{\phi}} N_{\mu}.
\end{equation}
We now use this solution for $J_{\mu}$ in (\ref{14a}) to arrive at the
field equation
\begin{equation}
\Box \phi-(2 \frac{\lambda^{2}}{l_{0}^{2}} \frac{\phi}{\bar{\phi}}
-\lambda\frac{\lambda^{2}}{l_{0}^{2}} \frac{\phi^{2}}{\bar{\phi^{2}}})\phi=0 
\label{18}\end{equation}
in which a linear and a quadratic term in $\phi$ appears in 
the dynamical mass-term.
We can get a more effective form of this equation if we
use in the linear term the back-ground value $\bar{\phi}$ for $\phi$, 
leading to
\begin{equation}
\Box \phi-
(2 \frac{\lambda^{2}}{l_{0}^{2}}-
\lambda\frac{\lambda^{2}}{l_{0}^{2}} \frac{\phi^{2}}{\bar{\phi^{2}}})\phi=0. 
\label{19}\end{equation}
This shows the effective contribution of the divergence theory 
(\ref{10}) to the dynamical mass-term in (\ref{11}). Equation
(\ref{19}) can be derived from the effective Lagrangian density
\begin{equation}
{\cal L}=\frac{1}{2l_{0}^{2}}
[\partial_{\mu}\phi\partial^{\mu}\phi-V(\phi)],~~~
V(\phi)=-2\frac{\lambda^{2}}{l_{0}^{2}}\phi^{2}+\frac{1}{2}
\lambda\frac{\lambda^{2}}{l_{0}^{2}} \frac{\phi^{4}}{\bar{\phi^{2}}}
\end{equation}
from which we can now define the ground state value $\phi_0$ of $\phi$ 
by minimising the potential $V(\phi)$. 
This gives the condition $\phi_{0}^{2}=\frac{2}{\lambda}\bar{\phi}^{2}$. 
Let us choose $\phi_{0}=(\frac{2}{\lambda})^{1/2}\bar{\phi}$ as
the ground state value  
on which to study the nature of physical excitations. 
The potential $V(\phi)$ can be expanded around $\phi_0$
to yield (neglecting constant terms)
\begin{equation}
V(\phi)= 8(\frac{\lambda}{l_{0}})^{2} 
(\phi-\phi_{0})^{2}+O((\phi-\phi_{0})^{3}),
\end{equation}
from which we infer that physical excitations of $\phi$ around 
the ground-state value $\phi_0$ provides 
a characteristic scale of mass of the order $\sim\frac{\lambda}{l_{0}}$.\\
We may therefore argue that in the broken phase of 
the Lorentz invariance an effective cosmological constant $\Lambda$ 
must appears which 
receives contributions proportional to $(\frac{\lambda}{l_{0}})^{2}$, namely
\begin{equation}
\Lambda \sim (\frac{\lambda}{l_{0}})^{2} 
\end{equation}
which in conjunction with (\ref{14}) yields
\begin{equation}
\Lambda \sim\frac{1}{R^{2}}.
\end{equation}
The agreement of this relation with 
the present observational bound for the cosmological constant is a 
remarkable consequence of the well-known empirical fact 
that the present universe
has just the characteristic size $R\sim 10^{29} cm$.\\

\vspace{1cm}
\begin{center}  
{\large {\bf  VI-Concluding remarks\\ }} 
\end{center}
\vspace{1cm}

In this note we have demonstrated that 
the broken phase of Lorentz invariance can provide a consistent
contribution to the cosmological constant via the mass
scale of an associated scalar condensate. 
The basic input was to consider the duality breaking as a cosmological 
effect.\\
We emphasize that there may be a potential dependence of 
the energy density of the vacuum on the entire mass spectrum
of physical fields in quantum field theory, for example the Higgs mass.
We have not commented on the dynamical reasons for why the 
corresponding contributions should cancel out. In this respect  
we must still look for more or less natural rules to
establish the applicability of the model presented for the prediction
of the value of the cosmological constant. We hope to address the issue
elsewhere another publication.\\     \\\\

{\bf References}\\\\
\begin{tabular}{r p{12cm}}
1. & Weinberg S  Rev. Mod. Phys. 61 (1989) 1\\
2. & Blokhintsev D I, Phys. Lett. 12 (1964), 272  \\
3. & Nielsen H B, Picek I, Nucl. Phys. B211 (1983), 269  \\
4. & Phillips P R, Phys. Rev. 139, 2B (1965), 491\\
5. & Salehi H, Work in progress\\
\end{tabular}

\end{document}